\documentclass[%
 reprint,longbibliography,preprintnumbers,
nofootinbib,
 amsmath,amssymb,
 aps,
pre,
]{revtex4-2}
\usepackage{graphicx}
\usepackage[utf8]{inputenc}
\usepackage{flushend}
\usepackage{dcolumn}
\usepackage{bm}
\usepackage{balance}

\usepackage{xcolor} 
\usepackage[normalem]{ulem} %

\usepackage[normalem]{ulem}

\usepackage{verbatim}
\usepackage{color,ulem}
\usepackage[english]{babel}

\def\be{\begin{equation}}
\def\ee{\end{equation}}
\def\bea{\begin{eqnarray}}
\def\eea{\end{eqnarray}}

\usepackage[utf8]{inputenc}

\input Starburst.fd
\newcommand*\initfamily{\usefont{U}{Starburst}{xl}{n}}\initfamily

\makeatletter 
    
\renewcommand\onecolumngrid{
\do@columngrid{one}{\@ne}%
\def\set@footnotewidth{\onecolumngrid}
\def\footnoterule{\kern-6pt\hrule width 1.5in\kern6pt}%
}

\renewcommand\twocolumngrid{
        \def\footnoterule{
        \dimen@\skip\footins\divide\dimen@\thr@@
        \kern-\dimen@\hrule width.5in\kern\dimen@}
        \do@columngrid{mlt}{\tw@}
}%

\makeatother

\newcommand{\beq}{\begin{eqnarray}}
\newcommand{\eeq}{\end{eqnarray}}
\usepackage{amsmath}
\usepackage{tikz}
\usetikzlibrary{decorations.pathmorphing}
\usetikzlibrary{shapes.misc}
\tikzset{cross/.style={cross out, draw=black, minimum size=8*(#1-\pgflinewidth), inner sep=0pt, outer sep=0pt},
cross/.default={1pt}}
\usetikzlibrary{patterns,math}
\usepackage[colorlinks = true,
            linkcolor = blue,
            urlcolor  = blue,
            citecolor =red,
            anchorcolor = blue]{hyperref}
            
\allowdisplaybreaks[4]

\begin{document}

\title{
The late-time attractor structure of dynamical black branes
}

 \author{Yan Liu$^{1,2}$}
\email{yanliu@buaa.edu.cn}
 \author{Hao-Tian Sun$^{2}$}
 \email{sunhaotian@buaa.edu.cn}
 
\affiliation{
\vspace{0.25cm}
$^1$Center for Gravitational Physics, Department of Space Science, 
Beihang University, Beijing 100191, China}
\affiliation{$^2$Peng Huanwu Collaborative Center for Research and Education, Beihang University, Beijing 100191, China}

\begin{abstract}
The ringdown phase of a perturbed black hole is conventionally described by a linear superposition of quasinormal modes. However, as the AdS  black brane approaches its final global equilibrium, this linear quasinormal mode description becomes inadequate, and nonlinear dynamics play a significant role in the late-stage evolution. We show that the interplay between nonlinear evolution and horizon dissipation in general relativity drives dynamical AdS black branes towards the final state along a unique path, independent of their initial perturbations. Through numerical simulations, we identify this late-time attractor and uncover the associated universal nonlinear behavior, characterized by a simple dimensionless relative amplitude in the dual hydrodynamic variables.   
\end{abstract}

\maketitle

\textit{Introduction.--} 
Black holes, as fundamental solutions of general relativity, play a central role in shaping our modern understanding of 
spacetime and gravity \cite{Barack:2018yly, Buoninfante:2024oxl, Berti:2025hly}. Through holographic  duality, their properties in 
anti-de Sitter (AdS) spacetime  are further connected to strongly coupled quantum field theories, making their behavior of broad theoretical interest \cite{Ammon:2015wua, Zaanen:2015oix, Hartnoll:2018xxg}. 

Near equilibrium, black hole dynamics are conventionally described by quasinormal modes (QNMs). 
The standard procedure for computing QNMs is well-known \cite{Kokkotas:1999bd, Nollert:1999ji, Berti:2009kk}. Starting from a black brane solution, 
one linearizes the Einstein equations around the equilibrium solution and performs a Fourier decomposition of the perturbations. Because the resulting equations are linear, modes with different spatial momentum 
$k$ decouple, and with proper physical boundary conditions at the horizon and at infinity one solves an eigenvalue problem for the complex frequency 
$\omega(k)$.  The solutions, which possess a nonzero imaginary part, define the quasinormal modes.

This approach, however, relies crucially on the validity of the linearized approximation.
The perturbative equations remain linear and decouple 
for different $k$  
only when the amplitude of each mode satisfies certain relations. 
However, during time evolution
the linearized solutions 
typically violate the consistency conditions required for linearization, exposing  contradictions that mark the breakdown of the linear regime. 

In this Letter, by evolving the full Einstein equations, we demonstrate that the linear description remains valid only for a finite time. As the system relaxes toward equilibrium, due to horizon dissipation, nonlinear effects inevitably dominate: they disrupt the  higher-order modes, and drive the AdS black brane along a  unique path to the final state.\\

\textit{Dissipation enhances nonlinear effects.--} 
We illustrate this mechanism with a simple example: 
a planer black brane where 
one spatial direction ($x$) is 
compactified. The periodicity of this direction sets a fundamental wavenumber determined by the system size, implying that all physical quantities along 
$x$ admit a discrete Fourier decomposition
in terms of sine or cosine function depending on the parity under $x\to -x$,  
\begin{align}
    g_{\mu\nu}(r,t,x)=\sum_{n=0}^{\infty}h_{\mu\nu} (r,t,k_n)\cos{k_nx}\,, \label{eq:decomp}
\end{align}
where $k_n=n\frac{2\pi}{L}$ and $L$ is the size of the compactified direction. 

Assume that the 
equations of motion 
contain both linear and nonlinear terms  
\begin{align}
   0= \mathcal{E}_{\mu\nu}[\nabla,g_{\mu\nu}]\supset g+g^2\,,
\end{align} 
where for illustration we keep only  
one linear and one nonlinear contribution. We  apply a Fourier decomposition to $g$, assuming that
$g$ is smooth and periodic along the spatial direction $x$, the equations of motion become 
\begin{align}
\begin{split}
    & \mathcal{E}_{\mu\nu}[\nabla,g_{\mu\nu}]\supset\sum_{s=0}^{\infty}h_s\cos{k_s x}+\left(\sum_{n=0}^{\infty}h_n\cos{k_n x}\right)^2\\
    &~~=\sum_{s=0}^{\infty}h_s\cos{k_s x}
    +\sum_{n,m=0}^{\infty}\frac{h_nh_m}{2}(\cos{k_{m+n}x}+\cos{k_{m-n}x})\,.
\end{split}
\end{align} 
This explicitly shows that the nonlinear term  couples different Fourier modes.  
The product $h_nh_m$ contributes to the equation for $h_s$ whenever $s=n+m$ or $s=|n-m|$. 

The nonlinear terms become negligible
only when the following condition holds: 
\begin{align}
    h_{m+n}\gg h_mh_n,\quad h_{m-n}\gg h_mh_n, \quad \forall n,m\in N. \label{eq:conditon}
\end{align} 
When this requirement is satisfied, one obtains a set of decoupled, linearized equations of motion, as in standard perturbation theory. 
However, due to the dissipative nature of the black brane horizon, the amplitude of each mode $h_n$ 
evolves differently in time. Consequently, the right-hand side of \eqref{eq:condition} could decay more slowly than the left-hand side. Hence, 
while condition \eqref{eq:condition} holds initially, it is often violated during the subsequent evolution.

Consider, for example, a system initially containing a 
fundamental mode 
$\{\omega_1,k_1\}$ and 
another mode $\{\omega_2,2k_1\}$, with comparable amplitude 
$h_1\sim h_2\ll1$. These two modes are decoupled in the beginning and each follows its own QNM evolution. Through nonlinear interactions, the fundamental mode 
sources a driven mode at momentum $2k_1$ and frequency $2\omega_1$ with an amplitude proportional to 
$h_1^2$.  
Although $h_1^2\ll h_2$ at early times, 
this sourced contribution remains negligible only if 
$2\,|\text{Im}[\omega_1]|>|\text{Im}[\omega_2]|$.
If instead $2\,|\text{Im}[\omega_1]|<|\text{Im}[\omega_2]|$,  
the nonlinear driven $2k_1$  mode decays more slowly than its 
linear counterpart (i.e. QNM with $\{\omega_2,2k_1\}$). 
 It therefore becomes the dominant contribution at late times, after the timescale  
\begin{align}
    t_{NL}=\frac{1}{2\,\text{Im}[\omega_1]-\text{Im}[\omega_2]}\log\frac{h_2}{h_1^2}\,.
\end{align}
Meanwhile, the fundamental mode always satisfies condition \eqref{eq:condition}, is well-approximated by a linear QNM, and ultimately together with its exciations governs the late-time evolution.

In general, whether a linear mode will be modified by nonlinear effects can be inferred from the linear spectrum. 
If the linear spectrum 
satisfies  
\begin{align}
\label{eq:condition}
    |n\mathrm{Im}[\omega_1]|<|\mathrm{Im}[\omega_n]|, \quad \exists\, n\in N^+
\end{align}
where  $\omega_1$ is fundamental mode and $\omega_n$ is $n$-th mode, then $n$-th mode will be altered by nonlinear effect within a finite time. 
This modification affects both its amplitude and its effective frequency.\\ 

\textit{
Unique attractor from hydrodynamic dispersion relation
--} For black branes in AdS, the linear QNMs for perturbation at large wave length are described by the hydrodynamics \cite{Kovtun:2005ev, Baier:2007ix, Bhattacharyya:2007vjd}. Near equilibrium, and under the decoupling assumption, hydrodynamic modes take the  form
\begin{align}
  \omega_n=\pm v_s k_n-i\Gamma k_n^2\,,~~~~\omega_n=-iD k_n^2\,, \label{eq:disp}
\end{align}
where $v_s$ is the speed of sound and $\Gamma$ and $D$ are the attenuation and diffusion constants. This dispersion relation
appears broadly in relativistic and nonrelativistic fluid dynamics \cite{Romatschke:2009im, Kovtun:2012rj},  holographic systems from Einstein–Maxwell-scalar theories and many other gravitational setups  \cite{Brigante:2007nu, Hartnoll:2018xxg} etc.

For our purpose, the precise values of the hydrodynamic coefficients are unimportant; what matters is the structure of the dispersion relation itself 
\cite{fnote1}. 
Combined with the condition \eqref{eq:condition}, it follows that nonlinear effects inevitably dominate after a finite time. Thus the system must undergo a nonlinear frequency shift after certain nonlinear time scale:
\begin{align}
    \omega_n\rightarrow n\omega_1\,.
    \label{eq:NLfrequency}
\end{align}

\begin{figure}[h!]
    \centering
    \includegraphics[width=0.45\textwidth]{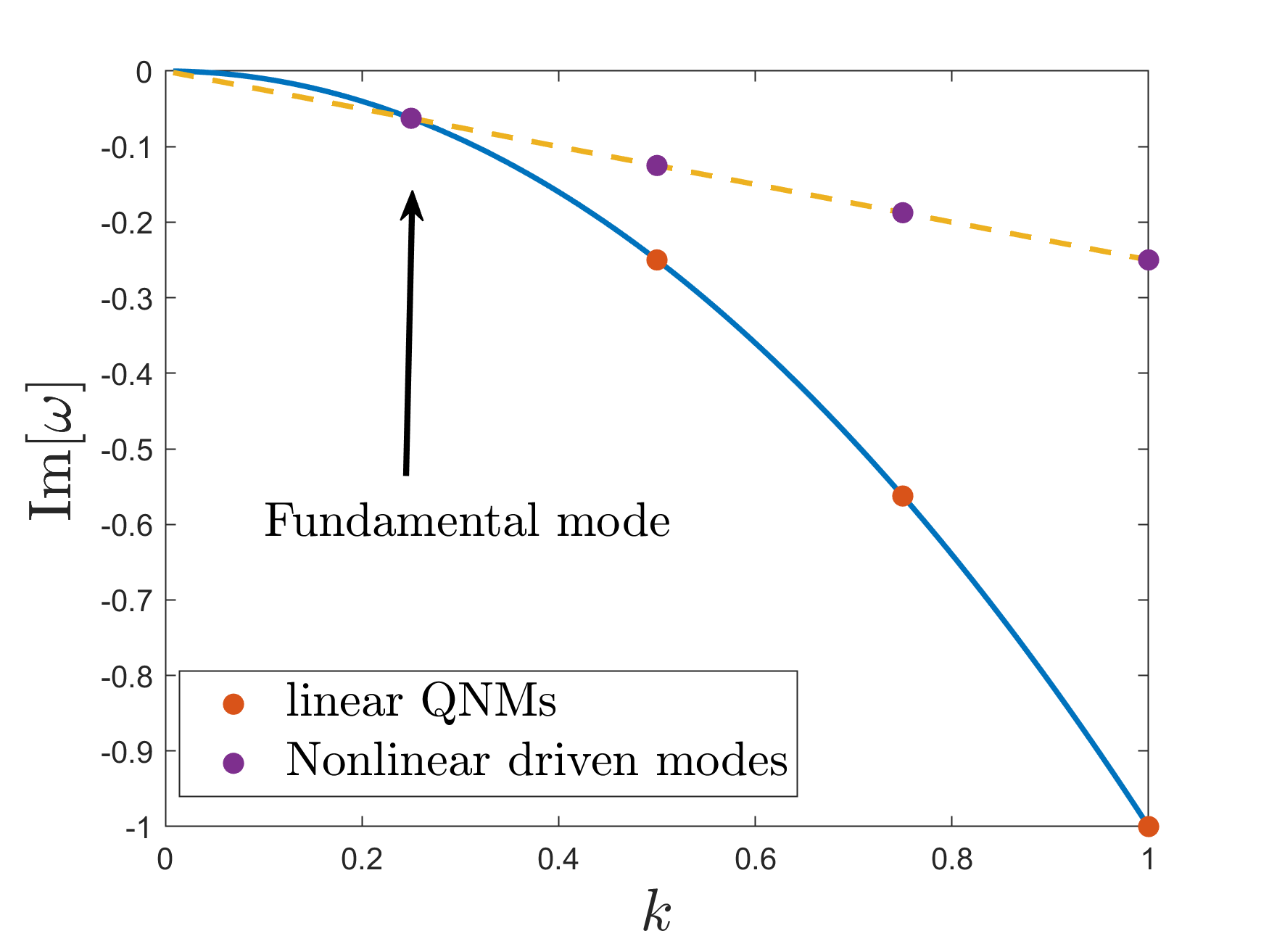}
    \caption{\small 
    This figure shows the typical spectrum of fluctuations for the system. In a compact system, the wavenumber $k$ can only take on  discrete values, as labeled by dots. The dispersion relation for  the hydro-modes \eqref{eq:disp} is plotted as red dots in the blue curve, which might play  important role in early-time evolution. 
    Late-time evolution is dominated by the fundamental mode and its excitation, shown as purple dots in the yellow dashed line. 
    }
    \label{fig:spectrum}
\end{figure}

Consequently, the spectrum of the system, defined as the set of complex frequencies associated with the Fourier amplitudes at each wave number, becomes time-dependent: it evolves according to the relative amplitudes of the modes rather than being fixed solely by the linear eigenfrequencies. In Fig. \ref{fig:spectrum}, we show two asymptotic spectra: a linear spectrum, valid when the amplitudes satisfy the condition \eqref{eq:conditon}; and the late-time driven spectrum that emerges once nonlinear effects dominate. 
Unlike linear QNMs, the amplitudes of the driven modes are not independent; they are fixed by nonlinear interactions. Their relative magnitudes  are determined intrinsically by the underlying theory and can only be obtained through a fully nonlinear analysis. Remarkably, both the late-time frequencies and amplitudes of the infinite driven modes are governed entirely by the fundamental mode.

This leads to a striking conclusion: 
the perturbed black brane's relaxation toward equilibrium follows a unique trajectory,  independent of the initial perturbation. We propose that a near-equilibrium attractor in the solution space arises from the interplay between the nonlinear dynamics of gravity and the dissipative nature of the horizon.\\

\textit{Attractor in Einstein gravity.--} To analyze the nonlinear behavior near equilibrium, we start from the four dimensional Einstein-Hilbert action,
\begin{equation}\label{action}
S\,=\frac{1}{2\kappa_N^2}\,\int d^4x \sqrt{-g}
\left[\mathcal{R}-2 \Lambda\right] 
\,,
\end{equation}
where $\mathcal{R}$ is the Ricci scalar, $\Lambda$ is the cosmological constant, and $\kappa_N^2=8\pi G_N$ is related to Newton's constant. For convenience, we set $\kappa_N^2=1$ and $\Lambda=-3$. 

We consider the following inhomogeneous ansatz for the metric in Eddington-Finkelstein (EF) coordinates \cite{chesler2014}:
\begin{align}
ds^2=&-2A(u,t,x)dt^2-\frac{2 \,du\, dt}{u^2}-2F(u,t,x)dtdx\notag\\
&+\Sigma(u,t,x)^2  \left(e^{B(u,t,x)}dx^2+e^{-B(u,t,x)}dy^2\right)\,,\label{eq:metric}
\end{align}
where the metric depends on the holographic radial coordinate $u$, the time $t$ and one spatial coordinate $x$. We assume that the remaining  spatial direction $y$ is homogeneous. The $x$-direction is taken to be periodic with period  $L$, which sets the fundamental wavenumber of the system. The components $\{g_{ty}, g_{xy}\}$ belong to a sector with odd parity under $y\to -y$, in contrast to the even-parity sector we focus on here. Therefore, if $\{g_{ty}, g_{xy}\}$ are set to zero initially, they remain zero throughout the evolution by parity symmetry. 

We verify the existence of a near-equilibrium attractor by evolving the system from different sets of initial data and checking whether the solutions converge to a unique late-time state. 
To drive the system away from equilibrium, we introduce inhomogeneous profiles for the boundary field-theory operators. According to the holographic dictionary, the boundary stress–energy tensor 
$\hat{T}_{\mu\nu}$ 	
  is determined by the near-boundary behavior of the bulk geometry.
 In (2+1)-dimensional field theory, it takes the form: 
\begin{align}
    \hat{T}_{\mu\nu}=\begin{bmatrix}
    E&J&0\\
    J&P+\sigma&0\\
    0&0&P-\sigma
    \end{bmatrix},
\end{align}
where $E$ is energy density, $J$ is energy current density, $P$ is pressure and $\sigma$ is shear stress. 

Initially, we set the energy density of boundary system to be homogeneous, $E(t=0,x)=E_0$, while imposing an inhomogeneous energy current of the form 
\begin{align}
    &J(t=0,x)=\sum_{n=1}^{\infty}j_n\sin(2\pi n x/L)\,.
\end{align}
Together with the initial profile $B(u,t=0,x)=0$ for the bulk metric function, other initial metric components are uniquely fixed by solving the constraint equations, thereby establishing all initial conditions.   
The choice of $j_n$ serves as an initial condition that determines the system's fundamental mode.   
Initially we can choose a finite set of modes with wavenumbers $n_1,\cdots,n_m$,  nonlinear interactions generate an infinite tower of additional modes whose momenta are integer linear combinations of the initial ones. Among these, the lowest nonzero wavenumber is $\text{gcd}(n_1,\cdots,n_m)$, which may differ from all initial wavenumbers. This establishes the fundamental mode of the system which 
ultimately governs the late-time evolution.

\begin{figure}[h!]
    \centering
    \includegraphics[width=0.485\textwidth]{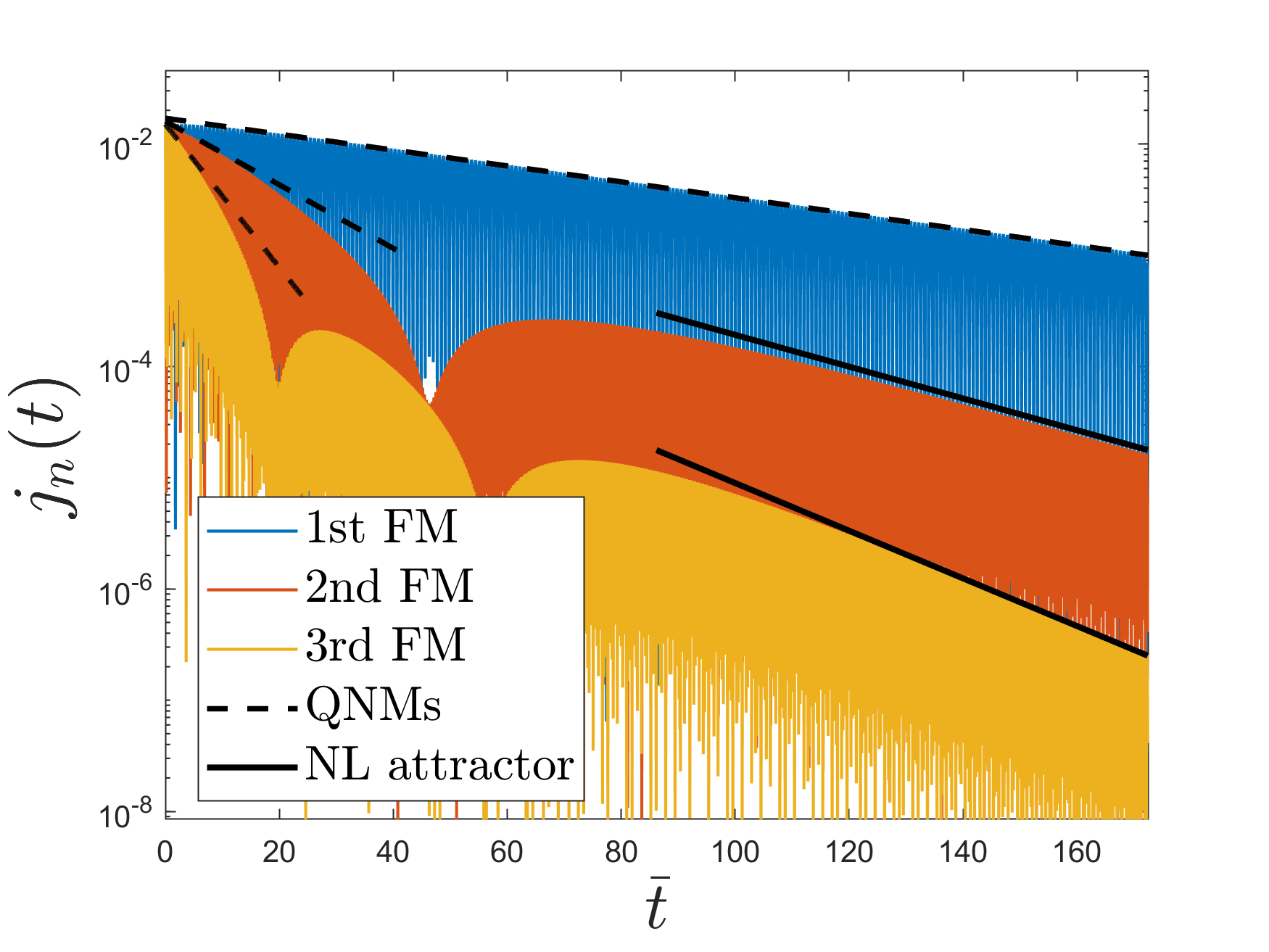}
    \caption{This figure shows the amplitude of first three Fourier modes (FM) for energy current as a function of $\bar{t}=t/L$.  Here $E_0=1$, $ L=400$.
  }
    \label{fig:QNM}
\end{figure}

We choose a set of initial $j_n$ with $\{j_1=j_2=j_3=j_4=1.5\times10^{-2},~j_{n>4}=0\}$, i.e. 
the first four modes $j_n$ have comparable magnitudes, the early-time evolution of each mode is governed by its corresponding QNMs, as shown in  Fig. \ref{fig:QNM} (dash black lines). However, once the nonlinear term become dominate,  the frequencies of the higher modes shift according to \eqref{eq:NLfrequency}. 
Their late time decay follows the driven mode, scaling as   $e^{2\text{Im}[\omega_1]t}$ and $e^{3\text{Im}[\omega_1]t}$ respectively. 
This behavior can also be seen in the solid black lines of Fig. \ref{fig:QNM}.

\begin{figure}[h!]
    \centering
    \includegraphics[width=0.48\textwidth]{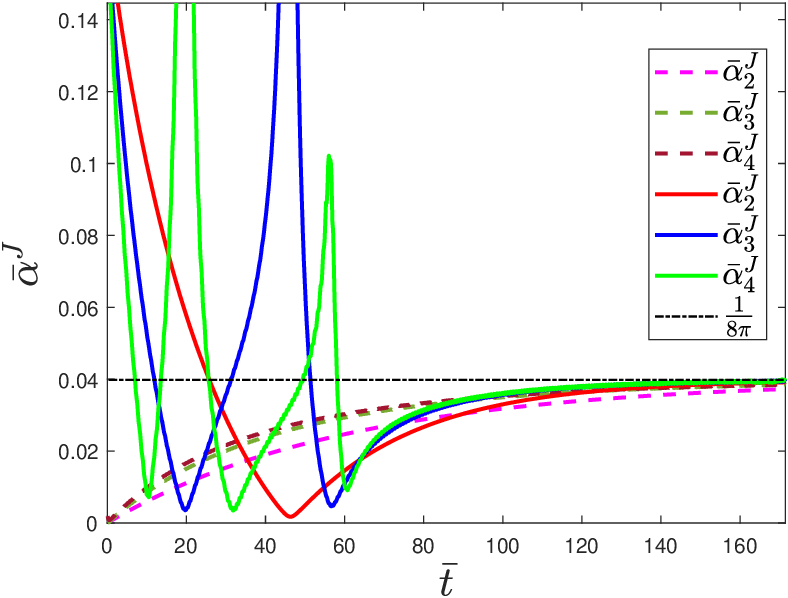}
    \caption{\small 
    The time evolution of the relative amplitude for the energy current. The dashed lines ( $L=800$, $E_0=1$) and solid lines
    ($L=400$, $E_0=1$) correspond to different initial conditions and different size. For all initial  conditions, ${\bar \alpha}_n$ approaches the same constant value, indicating that the final state is independent of the initial configuration.}
    \label{fig:alphaJ}
\end{figure}

The time evolution of $j_n$ for another set of initial condition with  $\{j_1=1.5\times10^{-2},~j_{n>1}=0\}$ is provided in the supplementary material, and it leads to the same conclusion as presented in the main text (dashed lines in Fig. \ref{fig:alphaJ}).  
For more generic initial condition with different sets of excitations, due to the nonlinearity of the system, 
all components of the metric in  \eqref{eq:metric} will be excited, and subsequently consequently evolve with a similar pattern at late times, as observed in quantities like energy density, entropy, pressure.
\\

\textit{Amplitude cascade of Fourier modes.--}
To illustrate the nonlinear effect on the amplitude,  we decompose a generic quantity $f(t,x)$ during the evolution as 
\begin{align}
    f(t,x)=f_0(t)+\sum_{n=1}^{\infty}f_n(t)\sin(2\pi n x/L)\,,
    \label{eq:struc}
\end{align} 
where 
$f(t,x)$ can represent any physical observable. 
We define the relative amplitude for different Fourier modes as
\begin{align}
\label{eq:ratio}
    \alpha_n=\frac{\tilde{f}_n(t)}{\tilde{f}_1(t) \tilde{f}_{n-1}(t)}\,,
\end{align}
where $\tilde{f}_n(t)$ is the envelope of $f_n(t)$, allowing us to focus on the amplitude. 

Our analysis shows that at late times, the leading order behavior of each mode is  $f_n(t)=f_ne^{-n\omega_1^it} \cos{n\omega^r_1 t}$ (up to a constant phase in the cosine function), so that $\tilde{f}_n(t)\approx f_ne^{-n\omega_1^it}$, where we have used that the linear QNM for the fundamental mode ($n=1$) is $\omega_1=\omega_1^r-i\omega_1^i$.  Consequently, the relative amplitude $\alpha_n$  asymptotically 
approaches a constant.

Fig. \ref{fig:alphaJ} shows the time evolution of the dimensionless quantity $\bar{\alpha}^J_n(t)=E_0^{2/3}\alpha^J_n/L$ 
for the energy current, where $E_0$ is the energy density of final black brane and $L$ is the system size. The corresponding behavior for energy density, $\alpha^E_n$, is provided in the supplementary material. While the early-time  evolution depends on the choices of initial conditions, all curves asymptotically approach a constant at late times. Notably, different modes ($n$)  converge to the same value of $\alpha$ within each quantity ( 
$\alpha^J$). 
At a late time, dimensionless $\bar{\alpha}^J_n$ satisfies  
\begin{align}
\label{eq:constant}
    \lim_{t\rightarrow\infty}\bar{\alpha}^J_n(t)=\frac{1}{8\pi}\,.
\end{align} 
The emergence of a constant late-time  $\bar{\alpha}^J$ indicates that the full dynamical  solution is attracted to a unique late-time state, at which all information about the initial configuration is erased. 

We therefore identify the late-time state as an  attractor solution. Due to the combined effects of nonlinearity and horizon dissipation, the physical quantity has a simple late-time form, 
\begin{align}
\label{eq:jnt}
    j_n(t)=(\alpha^J)^{n-1}j_1^n\sin (n\omega_1^rt)e^{-n\omega^i_1t}
\end{align}
where $n\in N$, $\omega_1^r=\text{Re}[\omega_1]$ and $\omega_1^i=-\text{Im}[\omega_1]$.

The relationship between the asymptotic values 
$\alpha^E$ and $\alpha^J$
  is determined by the conservation of the energy-momentum tensor, which gives 
\begin{align}
\alpha^E \approx v_s \alpha^J,
\end{align}
where 
$v_s=1/\sqrt{2}$
  is the speed of sound in the boundary theory. This relation is confirmed by our numerical results. We also confirmed numerically that $\alpha^P=2 \alpha^E$, consistent with 
  the underlying 
  conformal symmetry of the system. \\ 

\textit{Entropy and nonlinearty in the bulk.--} An important physical quantity related to the deep bulk geometry is the entropy, described by the area density of the apparent horizon  $\mathcal{S}=S/(4G)$. At different radial positions in the bulk, the area density is given by
\begin{align}
    S(u,t,x)=\Sigma(u,t,x)^2\,.
\end{align}
We find that the time evolution of area density $S(u,t,x)$ also follows the attractor solution at any radial position. At late times, the dimensionless quantity 
 $\bar{\alpha}^S(u)=\alpha^SE^{1/3}/L$ approaches a constant. 

\begin{figure}[h!]
    \centering
    \includegraphics[width=0.48\textwidth]{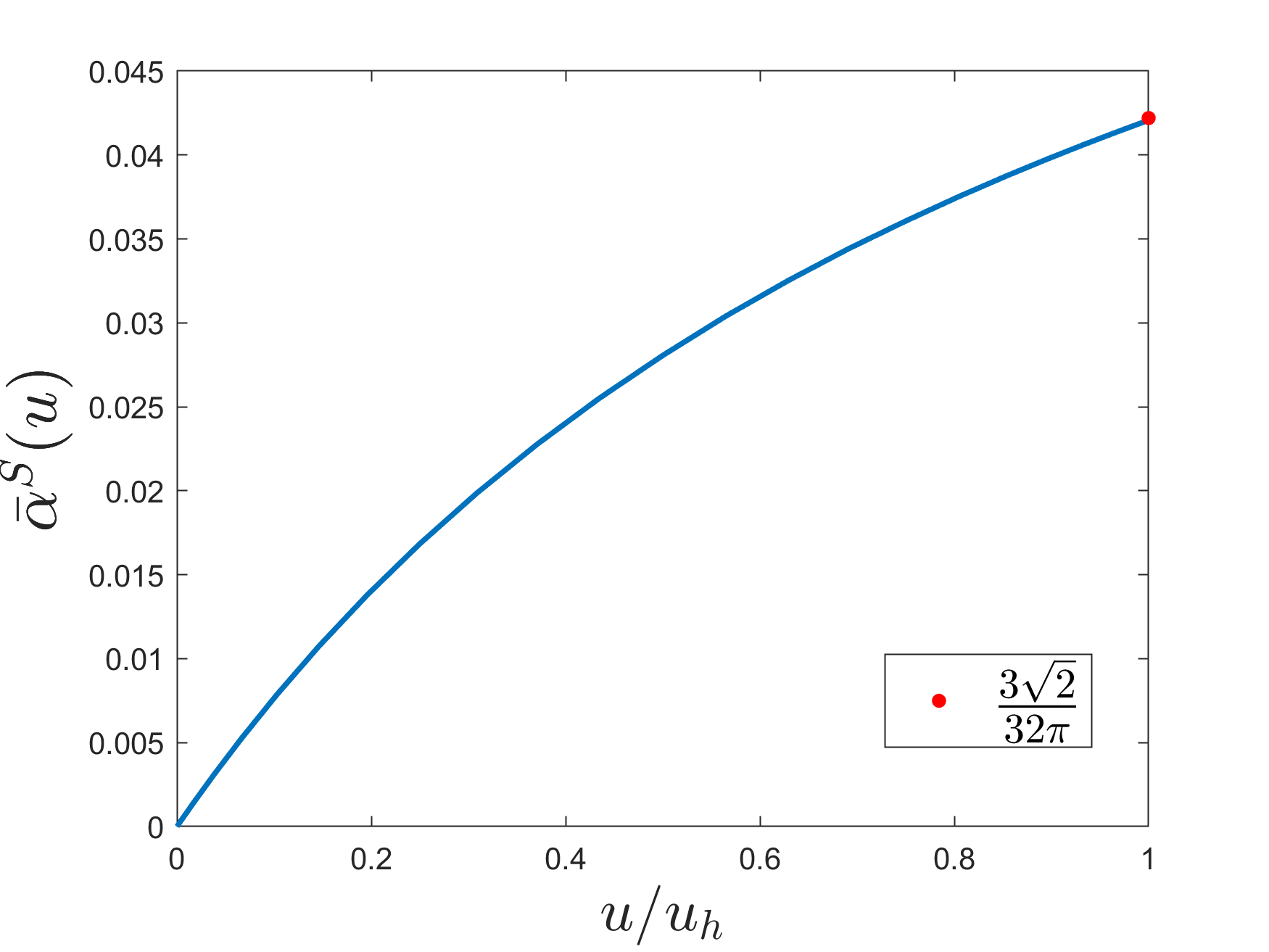}
    \caption{This figure shows that the final $\bar{\alpha}^S$ is a monotonic increasing function of radial position. 
   }
    \label{fig:alphaSu}
\end{figure}

The key difference from the boundary quantities is that 
$\bar{\alpha}^S(u)$ depends on the radial position $u$. This radial dependence provides a measure of  the nonlinearity in the area-density evolution. As shown in Fig.  \ref{fig:alphaSu},  $\bar{\alpha}^S(u)$ increases monotonically into the deep bulk, indicating that  
the nonlinear effects grow stronger 
from the boundary towards the horizon.

The horizon value $\bar{\alpha}^S(u_h)$  
can be related to $\bar{\alpha}^J$ via thermodynamics relations and  energy-momentum conservation. From the equilibrium thermodynamics of the final static black brane, 
\begin{align}
    &E=\frac{2}{3}T\mathcal{S}=\frac{4\pi}{3}TS\,,~~~~
    T=\frac{\mathcal{K}}{2\pi}=
    \frac{3}{4\pi}E^{1/3}\,,
\end{align}
where $\mathcal{K}$ is surface gravity of static black brane, we expect $\delta S=\frac{2}{3}E^{-1/3}\delta E$.  This leads to a simple relation between $\bar{\alpha}^E$ and  $\bar{\alpha}^S$,  allowing us to 
analytically determine the horizon value   
$\bar{\alpha}^S(u_h)$: 
\begin{align}
\label{eq:baras}
    \bar{\alpha}^S(u_h)&=\frac{3}{2}\bar{\alpha}^E
    =\frac{3}{2}v_s\bar{\alpha}^J=\frac{3\sqrt{2}}{32\pi}\,.
\end{align}
This analytically obtained value is marked by the red dot in 
Fig. \ref{fig:alphaSu}.

The simple attractor solutions 
\eqref{eq:jnt} for the energy density (pressure), energy current and entropy are either boundary quantities or directly related to them. 
In bulk, an attractor of the geometry  also exists; however, it generally no longer takes a simple closed form like \eqref{eq:jnt}. The $n$-th bulk spectrum with momentum 
$k_n$ becomes far richer during the late-time evolution, typically involving multiple oscillation frequencies with the decaying factor $e^{-n\omega^i_1t}$.  Nonetheless, the final late-time state remains unique.
\\

\textit{Discussion.--} 
We have shown that the combined effects of the intrinsic nonlinearity of gravity and horizon dissipation  drive the late-time dynamics of AdS black branes toward the final state along a unique path, independent of initial conditions. 
In Einstein gravity,  the dual hydrodynamic quantities evolve toward global equilibrium via a simple attractor structure given by \eqref{eq:struc} and \eqref{eq:jnt}. 
This structure is entirely governed by the fundamental mode and its nonlinear excitations, which is beyond the linear superposition of QNMs. Notably, the relative amplitudes between different modes $k_n$ converge to a universal constant given by  \eqref{eq:constant}. 

Furthermore, we find that the spectrum evolves over time, indicating that quasinormal modes describe the black hole spectrum only at a given stage of the evolution. This temporal change arises from the essential nonlinearity of the theory, which fundamentally alters the behavior of boundary excitations. 

The mechanism underlying the late-time attractor is likely generic in nonlinear dissipative dynamical  systems, motivating a systematic study of its universality. Future work could extend this analysis to higher-dimensional gravity theories  with or without matter fields, as well as to other nonlinear dissipative systems such as the Navier-Stokes equations. 
Furthermore, our preliminary results in Einstein-Maxwell theory confirm the universal value of \eqref{eq:constant} for the late-time nonlinearity ${\bar\alpha}$   \cite{longpaper}. 
It would be extremely interesting to understand better this universality from gauge/gravity duality.

\subsection*{Acknowledgments} 
 This work was supported by the National Natural Science Foundation of China Grants No. 12375041,  12447169 and 12575046. H-T. S also acknowledges the support from the Postdoctoral Fellowship Program of CPSF under Grant No. GZC20252777.

\bibliographystyle{apsrev4-2.bst}
\bibliography{ref}
\newpage
\onecolumngrid
\appendix 
\clearpage
\renewcommand\thefigure{S\arabic{figure}}    
\setcounter{figure}{0} 
\renewcommand{\theequation}{S\arabic{equation}}
\setcounter{equation}{0}
\renewcommand{\thesubsection}{SI\arabic{subsection}}

\section*{Supplementary Information}

\subsection{Equations of motion and boundary conditions}
We consider a four-dimensional holographic setup described by the action: 
\begin{equation}\label{action}
S\,=\frac{1}{2\kappa_N^2}\,\int d^4x \sqrt{-g}
\,\left[\mathcal{R}-2 \Lambda\right] + \text{bdy terms}\,,
\end{equation}
where $\kappa_N^2=8\pi G_N$ is the 4D gravitational Newton’s constant,  $\Lambda=-3$ is the negative cosmological constant. 
Boundary terms are included to ensure a well‑defined variational principle.  

For the ansatz \eqref{eq:metric},
the asymptotic AdS boundary is located at $u=0$.
Rather than using the event horizon -- which depends on the complete spacetime history -- we adopt the apparent horizon at $u=u_A$ as our interior boundary in non-equilibrium scenarios. The apparent horizon corresponds to the outermost trapped null surface that forms behind the event horizon.  

We use the formalism developed by Chesler and Yaffe to solve the problem \cite{chesler2014}. The bulk equations of motion 
are as follows: 
\begin{align}
u\Sigma{}{''} +2\Sigma{}{'}+\frac{{u B{}{'}}^2 }{4}&\Sigma=0\,,\label{eqS}\\
F''+\frac{2- B 'u}{u}F'+&\left( \frac{B '^2}{2}- B ''-\frac{2 \Sigma '^2}{ \Sigma ^2}-\frac{2 B '( \Sigma + \Sigma 'u)}{ \Sigma u}\right)F=-\frac{-2\partial_x \Sigma '+2 B '\partial_x \Sigma + \Sigma \partial_x B' - B '\partial_x B  \Sigma }{ \Sigma u^2}-\frac{2 \Sigma '\partial_x \Sigma }{ \Sigma ^2u^2}\,, \label{eqF}\\
{d_+\Sigma }'+\frac{ \Sigma'{d_+\Sigma }}{\Sigma }=
&\frac{e^{-B}}{8 \Sigma ^3 u^2} \Big(-2 F \Sigma ^2 u^2 \left(u^2 B' F'-B_x'\right)+2 B_x \Sigma ^2 u^2 \left(F'-F B'\right)+2 F_x \Sigma ^2 u^2 B'-F^2 \Sigma ^2 u^4 \left(B'\right)^2-12 e^B \Sigma ^4\notag\\
&-2 B_{xx} \Sigma ^2-4 \Sigma  \left(-\Sigma _{xx}+u^2 \Sigma ' \left(-B_x F+F_x+2 F^2 u-F u^2 F'\right)+B_x \Sigma _x+F u^2 \Sigma _x'+F^2 u^4 \Sigma ''\right)\notag\\
&+2 B_x^2 \Sigma ^2-2 \Sigma ^2 u^2 F_x'-4 \Sigma _x \left(\Sigma _x-F u^2 \Sigma '\right)+\Sigma ^2 u^4 \left(F'\right)^2\Big)
\,, 
\label{eqdpS}\\
d_+B'+\frac{d_+B \Sigma '}{\Sigma }=&-\frac{e^{-B}}{4 \Sigma ^4} \Big{(}4 e^B d_+\Sigma  \Sigma ^3 B'+F \left(-2 \Sigma  B' \left(B_x \Sigma -2 \Sigma _x+\Sigma  u^2 F'\right)+2 \Sigma ^2 B_x'-4 \Sigma  \Sigma _x'+4 \Sigma _x \Sigma '+4 \Sigma  u^2 F' \Sigma '\right)\notag\\
&+2 F^2 u \left(\Sigma  u \left(2 \Sigma ''-\Sigma  B''\right)+\Sigma ^2 u \left(B'\right)^2-2 \Sigma  B' \left(\Sigma +u \Sigma '\right)+4 \Sigma  \Sigma '-2 u \left(\Sigma '\right)^2\right)+\Sigma ^2 \left(2 F_x'-u^2 \left(F'\right)^2\right)\notag\\
&-4 \Sigma _x \Sigma  F'\Big{)}
\,, 
\label{eqdpB}\\
\frac{u A''}{2}+A'=&-\frac{e^{-B}}{4 \Sigma ^4 u^3} \Big(-e^B d_+B \Sigma ^4 u^2 B'+2 F \Sigma  u^2 \left(-\Sigma _x B'+F u \left(\Sigma ' \left(u B'-2\right)-u \Sigma ''\right)+\Sigma _x'\right)\notag\\
&+\Sigma ^2 u^2 \left(F \left(B_x B'-B_x'\right)+F^2 u \left(u B''-u \left(B'\right)^2+2 B'\right)+B_x F'-F_x'+u^2 \left(F'\right)^2\right)\notag\\
&-4 e^B \Sigma ^3 u^2 d_+\Sigma '-6 e^B \Sigma ^4+2 F u^2 \Sigma ' \left(F u^2 \Sigma '-\Sigma _x\right)\Big)
\,,
\label{eqA}
\end{align}
together with 
\begin{align}
&2 F^2 u^2 \left(\Sigma  B' \left(3 B_x \Sigma +4 \Sigma _x+3 \Sigma  u^2 F'\right)-2 \Sigma ^2 B_x'+4 B_x \Sigma  \Sigma '-4 \Sigma  \Sigma _x'+8 \Sigma _x \Sigma '+6 \Sigma  u^2 F' \Sigma '\right)\notag\\
&+2 F^3 u^3 \left(\Sigma ^2 u B''-2 \Sigma ^2 u \left(B'\right)^2+2 \Sigma  B' \left(\Sigma -2 u \Sigma '\right)-4 u \left(\Sigma '\right)^2\right)+4 e^B F \Sigma ^3 u^2 \left(A u \left(2 \Sigma ' \left(u B'-1\right)-u \Sigma ''\right)-\dot{B} \Sigma '+\dot{\Sigma }'\right)\notag\\
&
-2 e^B \Sigma ^4 \left(B_x \left(\dot{B}-2 A u^2 B'\right)+u^2 \left(2 A B_x'-\left(B' (\dot{F}-2 A_x)\right)+\dot{F}'-2 A_x'\right)-\dot{B}_x\right)\notag\\
&+4 e^B \Sigma ^3 \left(\Sigma _x \left(\dot{B}-2 A u^2 B'\right)+u^2 \left(2 A \Sigma _x'+\dot{F} \Sigma '-\dot{\Sigma } F'\right)-\dot{\Sigma }_x\right)+2 \Sigma ^2 \left(2 e^B \Sigma _x \left(\dot{\Sigma }-2 A u^2 \Sigma '\right)+B_x F_x\right)\notag\\
&+4 F_x \Sigma _x \Sigma
+e^B F \Sigma ^4 u^2 \left(-2 \left(2 u^2 \left(A''-A B''\right)+\dot{B}'\right)+4 u A' \left(u B'-2\right)+2 B' (4 A u+\dot{B})-3 A u^2 \left(B'\right)^2\right)\notag\\
&-F \Sigma ^2 \left(4 u^2 \left(e^B \Sigma ' \left(\dot{\Sigma }-2 A u^2 \Sigma '\right)+F_x B'\right)-2 B_{xx}+3 B_x^2+4 B_x u^2 F'\right)\notag\\
&-4 F \Sigma  \left(-\Sigma _{xx}+2 B_x \Sigma _x+2 u^2 \left(F_x \Sigma '+\Sigma _x F'\right)\right)-8 \Sigma _x^2 F=0\,,\label{eqcont}
\end{align}
and 
\begin{align}
&-2 F^2 \Sigma ^2 u^4 A''+2 \Sigma ^2 u^2 A' \left(F^2 u^2 B'+2 e^B \dot{\Sigma } \Sigma -B_x F+F_x-2 F^2 u\right)
+2 A F^2 \Sigma ^2 u^4 B''-2 A_x \Sigma ^2 u^2 F'+2 A_x B_x \Sigma ^2\notag\\
&+2 \Sigma  u^2 B' \left(A \left(e^B (-\dot{B}) \Sigma ^3+B_x F \Sigma -2 \Sigma _x F+2 F^2 \Sigma  u\right)+2 A F^2 u^2 \Sigma '+F \Sigma  (\dot{F}-A_x)\right)-2 A F^2 \Sigma ^2 u^4 \left(B'\right)^2+4 F \Sigma ^2 u^2 A_x'\notag\\
&-8 A e^B \Sigma ^3 u^2 \dot{\Sigma }'-2 A F \Sigma ^2 u^2 B_x'+2 A B_x \Sigma ^2 u^2 F'-2 A \Sigma ^2 u^2 F_x'+4 A F \Sigma  u^2 \Sigma _x'-4 A \Sigma _x F u^2 \Sigma '-2 \dot{F} B_x \Sigma ^2-2 F \Sigma ^2 u^2 \dot{F}'\notag\\
&-4 A F^2 \Sigma  u^4 \Sigma ''+4 A F^2 u^4 \left(\Sigma '\right)^2-8 A F^2 \Sigma  u^3 \Sigma '+4 e^B \ddot{\Sigma } \Sigma ^3-4 e^B \dot{A} \Sigma ^3 u^2 \Sigma '+e^B \dot{B}^2 \Sigma ^4-2 A_{xx} \Sigma ^2+2 \dot{F}_x \Sigma ^2=0\,.\label{eq:cons2order}
\end{align}
In the expressions above, we have introduced the directional derivative 
$d_+\mathcal{F}:= \dot{\mathcal{F}}-\frac{u^2 A}{2}\mathcal{F}{'}$, where the prime ($'$)  denotes differentiation with respect to the radial coordinate $u$,  and the dot ($\dot{}$) denotes the time derivative $\partial_t$. Additionally,  $\mathcal{F}_x\equiv \partial_x \mathcal{F}$ denotes the spatial derivative with respect to $x$,  $\mathcal{F}_{xx}\equiv \partial_x^2 \mathcal{F}$ is the second-order spatial derivative with  respect to $x$. For our time evolution scheme, we will solve the system using~\eqref{eqS} to~\eqref{eqA}. We will check that the last two equations~\eqref{eqcont} and~\eqref{eq:cons2order} are satisfied in our numerics. 
The detailed evolution scheme can be found in
\cite{LiLi:2025swx}. 

\subsection{Analysis of the second set of initial conditions}

\begin{figure}[h!]
    \centering
    \includegraphics[width=0.48\textwidth]{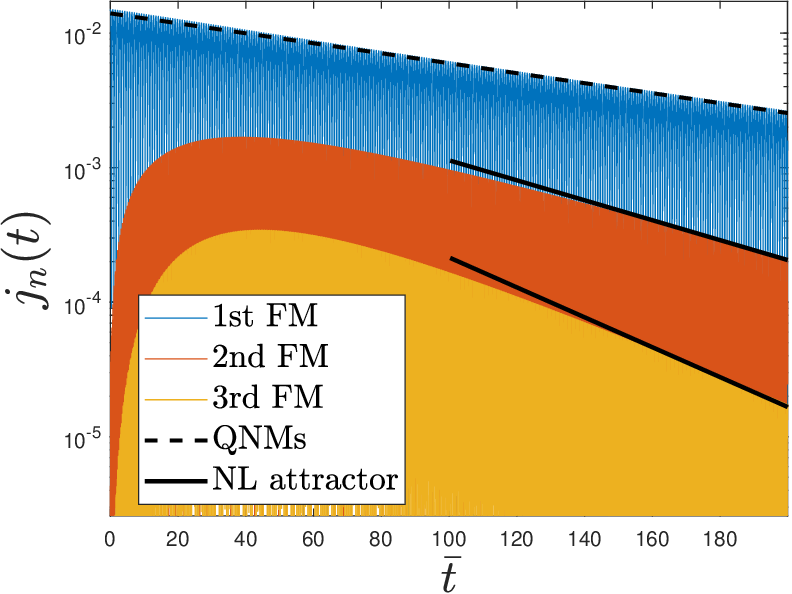}
    \caption{\small 
    The time evolution of the amplitude for energy current $j_n(t)$ in a log plot from initial condition ${j_1=10^{-2},j_{n>1}=0}$. Here $E_0=1$, $ L=800$.
    Due to nonlinear effects, Fourier modes $j_2$ and $j_3$ are excited during the time evolution. Their late time decay is also governed by the driven mode, scaling as  $e^{2\text{Im}[\omega_1]t}$ and $e^{3\text{Im}[\omega_1]t}$ respectively, as shown in the black line. }
    \label{fig:alphaJ2}
\end{figure}

For another initial condition,  ${j_1=1.5\times10^{-2},j_{n>1}=0}$, the evolution of the amplitude for energy current $j_n(t)$ is shown in Fig. \ref{fig:alphaJ2}. At early times, higher wavenumber modes are excited due to nonlinear effects. Since initially the condition \eqref{eq:conditon} is violated, the QNM region for higher wavernumber modes disappears. 
At late times, these modes also evolve into the unique attractor solution, same as the results from the initial condition shown in the main text. These excitations go beyond the linear superposition of QNMs, 
reinforcing the notion that 
QNM is not a complete basis \cite{Besson:2024adi}.  
This supports our proposal of a late-time unique attractor for the dynamical black brane.

\subsection{Comment on the relative amplitude of the Fourier modes}
For parity even sector,
we have 
\bea
E&=&E_0+\sum_{n=1}^\infty E_1^n (\alpha^E)^{n-1}\,\sin (n \omega_1^r t) e^{-n\omega_1^i t}\cos(n k x)\,,\\
J&=&\sum_{n=1}^\infty J_1^n (\alpha^J)^{n-1}\,\cos (n \omega_1^r t) e^{-n\omega_1^i t}\sin(n k x)\,,
\eea
where 
\be
k=\frac{2\pi}{L}\,,~~~\omega_1=\omega_1^r-i \omega_1^i=v_s k -i \Gamma  k^2\,.
\ee

From $\partial_t E+\partial_x J=0$, we have 
\be
E_1^n(\alpha^E)^{n-1}\omega_1^r=kJ_1^n(\alpha^J)^{n-1}\,,~~~ n\in N^+
\ee
where we have used the fact that when $L$ is large enough, the relation $\omega_1^r\gg \omega_1^i$ is satisfied.  

Therefore
\be\label{eq:ratio-alphaEJ}
\alpha^E=\alpha^J v_s\,.
\ee

\begin{figure}[h!]
    \centering
   \includegraphics[width=0.45\textwidth
   ]{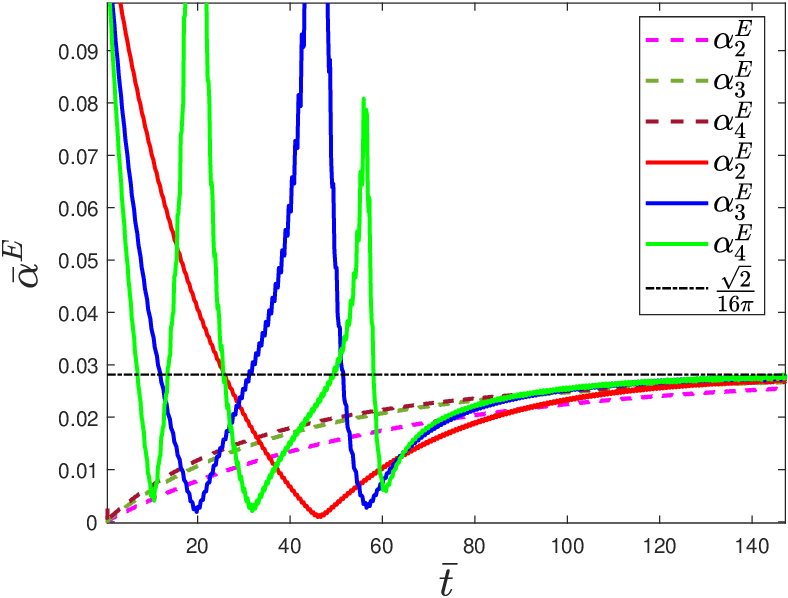}~~~~~~
   \includegraphics[width=0.45\textwidth
   ]{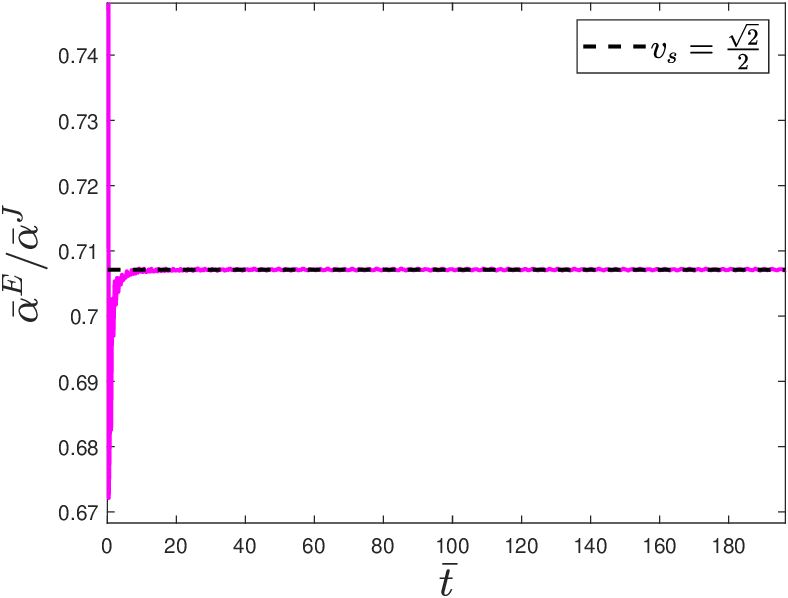}
    \caption{The time evolution of relative amplitude ${\bar{\alpha}^E}$ for energy density ({\em left}) and the ratio ${\bar{\alpha}^E}/{\bar{\alpha}^J}$ ({\em right}). 
    }
    \label{fig:alphaE}
\end{figure}

The left panel in Fig. \ref{fig:alphaE} shows the evolution of the amplitude from Fourier modes in energy density  $\alpha^E_n$. The right panel in Fig. \ref{fig:alphaE} confirmed the above relation \eqref{eq:ratio-alphaEJ}.

From the conservation equation 
\be
\partial_t J+\partial_x P-\partial_x\sigma=0
\ee
and the relation $P=E/2$, together with $\sigma\ll P$ we have 
\be
(\alpha^J)^{n-1}J_1\omega_1^r=\frac{k}{2}E_1^n(\alpha^E)^{n-1}\,,~~~n\in N^+ 
\ee
which is consistent with the relation \eqref{eq:ratio-alphaEJ}.

\begin{figure}[h!]
    \centering
   \includegraphics[width=0.45\textwidth
   ]{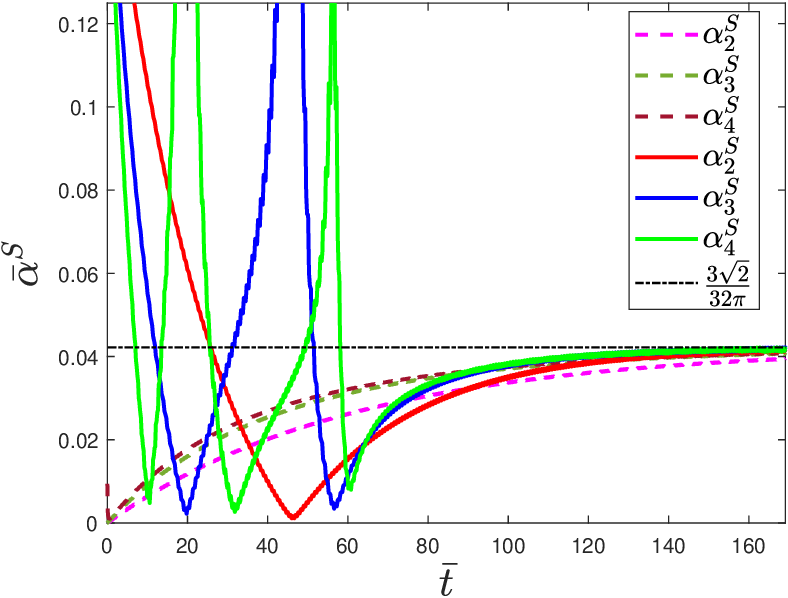}~~~~~~
   \includegraphics[width=0.45\textwidth
   ]{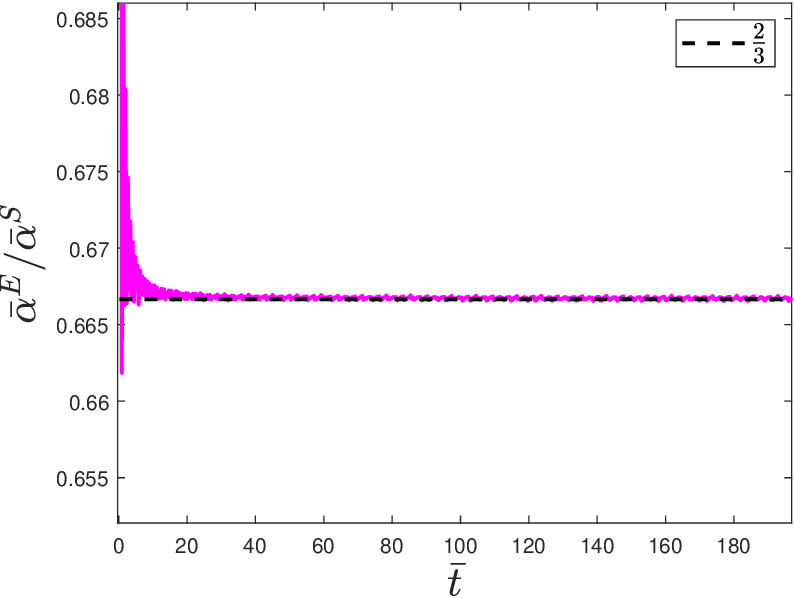}
    \caption{The time evolution of relative amplitude ${\bar{\alpha}^S}$ for area density at apparent horizon ({\em left}) and the ratio ${\bar{\alpha}^E/\bar{\alpha}^S}$ ({\em right}).
    }
    \label{fig:alphaS}
\end{figure}

Let us briefly comment on the relation \eqref{eq:baras}. For the horizon area which is related to the entropy, we have 
\be
S=S_0+\sum_{n=1}^\infty S_1^n (\alpha^S)^{n-1}\,\sin (n \omega_1^r t) e^{-n\omega_1^i t}\cos(n k x)\,,
\ee
From the relation $\delta S=\frac{2}{3}E^{-1/3}\delta E$ where $\delta S=S-S_0, \delta E=E-E_0$, we have 
\be
S_1^n (\alpha^S)^{n-1}=\frac{2}{3}
E_1^n (\alpha^E)^{n-1}\,, ~~~n\in N^+\,.
\ee
Therefore
\be
\alpha^S=\frac{3}{2}\alpha^E E_0^{1/3}\,,~~~~ S_1=\frac{2}{3}E_1\frac{1}{E_0^{1/3}}\,.
\ee
These two relations have been checked numerically in Fig. \ref{fig:alphaS}. 

Among the relative amplitudes, $\alpha^J, \alpha^E, \alpha^P, \alpha^S$, only one is independent. It should be very interesting to analytical calculate \eqref{eq:constant} from holography. Its value might be related to the calculation on three-point function in e.g. \cite{Pantelidou:2022ftm, Pan:2024bon}. 
\\

\begin{figure}[h]
    \centering
    \includegraphics[width=0.5\textwidth]{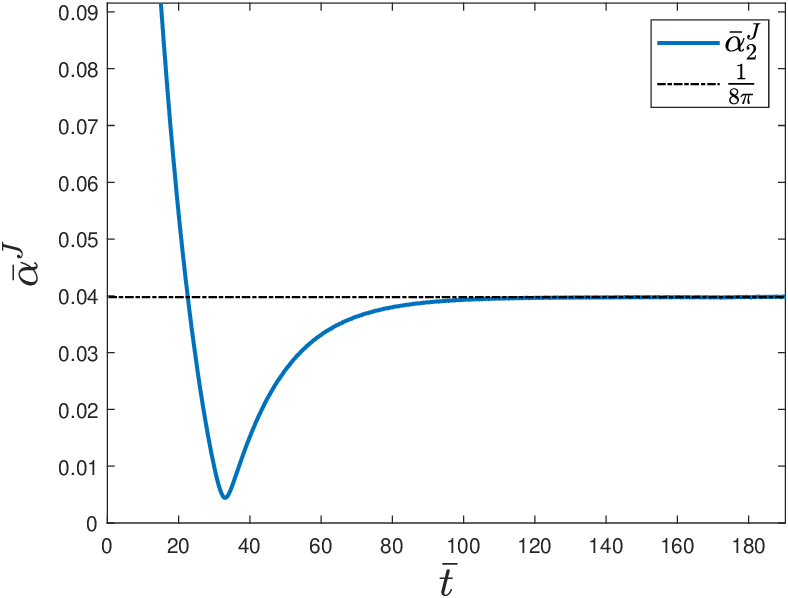}
    \caption{The time evolution of the relative amplitude  ${\bar{\alpha}^J}$. Here  $L=200$, $E_0=1$. Note that we choose a smaller $L$ here  than in  the main text, so that  ${\bar{\alpha}^J}$ reaches the saturation value  $\frac{1}{8\pi}$ more quickly.  This figure  shows that the relation  $\bar{\alpha}^J=\frac{1}{8\pi}$ is well satisfied at late times.}
    \label{fig:my_label}
\end{figure}

Finally, given that the ratio in Eq. \eqref{eq:ratio} approaches a constant at late times, as confirmed in Fig. \ref{fig:alphaJ},  Fig. \ref{fig:alphaE} and Fig. \ref{fig:my_label}, we can summarize all the higher order terms. 
The summation gives 
\bea
E(t,x)&=&E_0+\frac{E_1 e^{-\omega_1^i t}}{2}\bigg[
\frac{\sin(\omega_1^r t+k x)}{1+ E_1^2(\alpha^E)^2 e^{-2 \omega_1^i t}  -2E_1\alpha^E e^{-\omega_1^i t} \cos(\omega_1^r t+k x))}
+(x\to -x)
\bigg]
\,
,\\
J(t,x)&=&\frac{J_1 e^{-\omega_1^i t}}{2}\bigg[
\frac{\sin(\omega_1^r t+k x)}{1+J_1^2(\alpha^J)^2 e^{-2\omega_1^i t} -2 J_1\alpha^J e^{-\omega_1^i t} \cos(\omega_1^r t+k x))}-(x\to -x)
\bigg]
\,.
\eea

It can be further simplified under the assumption $t\to \infty$ as 
\bea
E&\simeq&E_0+E_1 e^{-\omega_1^i t}
\frac{\sin(\omega_1^r t)\cos(k x)}{1-4 E_1\alpha^E e^{-\omega_1^i t} \cos(\omega_1^r t)\cos(k x))}\,,\\
J&\simeq&J_1 e^{-\omega_1^i t}
\frac{\cos(\omega_1^r t)\sin(k x)}{1-4 J_1\alpha^J e^{-\omega_1^i t} \cos(\omega_1^r t)\cos(k x)}\,,
\eea
which 
satisfies a generalized Burgers equation
\be
\partial_t J =\omega_1^i \partial_x^2 J+8 \alpha^J k  J\partial_x J-\omega_1^r\tan(\omega_1^r t)( J+4 \alpha^J \cot(k x)J^2)\,.
\ee
The last term plays a role of stochastic noise and the above equation contains the important nonlinear term $ff_x$ which is similar to NS equation. 
It should be interesting to explore the relation  to (non-linear) hydro equations. It would be interesting to reproduce these hydrodynamics structures from holography.

\subsection{Comment on the dimensionless constants}
To construct a dimensionless constant $\alpha$ from the final-state quantities $T$, $E_0$, $S$, and $L$, there are infinitely many possibilities. 
If we only consider scaling symmetry, the $\alpha^J$ can be a dimensionless quantity like 
\begin{align}    
\frac{\alpha^J}{L^3},\ \alpha^J E_0,\ \alpha^J T^3,\ \frac{\alpha^J S}{L},\cdots\,. \end{align}

However, among these infinite dimensionless quantities, only certain combinations can characterize the dynamics of the evolution. 
Specifically, for different values of $E_0$ (where $S$ and $T$ are degenerate with the energy density $E_0$) and $L$, the evolution should yield the same value of $\bar{\alpha}^J$.
In Einstein gravity, the physically  meaningful dimensionless combination $\bar{\alpha}^J$, as shown in the main text, is 
\begin{align}
    \bar{\alpha}^J=\frac{\alpha^J E_0^{2/3}}{L}\,.
\end{align}
Here $E_0$ can be replaced by $T$ or $S$ due to their degeneracy in Einstein gravity. 
While using a different combination would alter the numerical constant, the descriptions remain equivalent.

In Einstein Maxwell theory, where $E$ and $T$ are no longer degenerate, our primarily results \cite{longpaper} indicate that the meaningful combination is 
\begin{align}
    \bar{\alpha}^J=&\frac{\alpha^J TE_0^{1/3}}{L}=\frac{3}{4\pi}\frac{1}{8\pi}\,.
\end{align}

\end{document}